\documentclass{article}
\usepackage{arxiv}
\usepackage[colorlinks,allcolors=blue]{hyperref}
\usepackage[normalem]{ulem}
\hypersetup{colorlinks,urlcolor=blue}

\usepackage{tabularx}
\usepackage{multirow}
\usepackage[table,xcdraw]{xcolor}
\usepackage{longtable}
\usepackage{graphics}
\usepackage{graphicx}
\title{Multimodal Group Activity Dataset for Classroom Engagement Level Prediction} % Article title

\author{
    Alpay Sabuncuoğlu \footnote{Corresponding Author}\\
    UNVEST R\&D Center\\
    \texttt{asabuncuoglu13@ku.edu.tr} \\
    \And
    T. Metin Sezgin\\
    Koc University\\
    \texttt{mtsezgin@ku.edu.tr} \\
}

\begin{document}
\maketitle

\begin{abstract}
We collected a new dataset that includes approximately eight hours of audiovisual recordings of a group of students and their self-evaluation scores for classroom engagement. The dataset and data analysis scripts are available on our open-source repository. We developed baseline face-based and group-activity-based image and video recognition models. Our image models yield 45-85\% test accuracy with face-area inputs on person-based classification task. Our video models achieved up to 71\% test accuracy on group-level prediction using group activity video inputs. In this technical report, we shared the details of our end-to-end human-centered engagement analysis pipeline from data collection to model development.
\end{abstract}

\section{Notes for Practitioners}

\begin{itemize}
    \item We shared our dataset, preprocessing steps and baseline analysis code, open-source at \url{https://github.com/asabuncuoglu13/classroom-engagement-dataset}. Practitioners can download our dataset using the script provided in this repository. Use the links in Table \ref{table:statistics} to quickly explore a sample session.
    \item The dataset contains 26540 frames with self-engagement evaluation scores. We utilized recent state-of-the-art deep learning models to extract features from this audiovisual data. We also released the resulted OpenFace, and OpenPose vectors for the use of other researchers to decrease the carbon footprint in the replication process (see Table \ref{table:statistics})
    \item Our MobileNet-MaxOut face-based prediction image model achieved up to 85\% test accuracy and MoViNetA4-powered video prediction model yielded 68\% test accuracy. We released the Jupyter notebooks in our Github repo.  
    \item We created an interactive dashboard to present the model results in a student-centric format. We detailly explained the research and development process on \cite{sabuncuoglu2023dashboard}.
    \item Overall, this research can accelerate the adaptation of recent deep learning models in exploring interpretable features to improve the current state of 21st-century learning environments.
    
\end{itemize}

\section{What is Classroom Engagement?}

Classroom engagement is a multi-component term that coins active involvement in learning, discussion, and reflection with peers, teachers, and materials in a classroom environment with three dimensions: Affective engagement, behavioral engagement, and cognitive engagement \cite{fredricks_school_2004}. Evaluating engagement is a challenging task, considering this multi-component definition of the term. Two methods, independent observers and self-evaluation, have been used in previous studies to measure engagement levels. The observational methods can be performed in the classroom in real-time or can be assessed via watching recorded videos. All these methods have their advantages and disadvantages. For example, observational methods require human effort to train observers and check annotation reliably. On the other hand, self-evaluation can produce unreliable results from social desirability bias and memory recall limitations \cite{dmello_advanced_2017, nan_2021}.

Existing work on engagement recognition presents only a limited amount of audiovisual dataset, which restricts the accurate representation of an engaging classroom environment. The need for a comprehensive and publicly-available dataset is acknowledged, yet the effort in this area needs to be improved \cite {dewan_engagement_2019}. Most of the video datasets have been collected online, where students interact with an interface while a camera faces directly toward the participants \cite{delgado_student_2021, sottilare_measuring_2021}. Datasets like DAiSEE \cite{gupta_daisee_2022} only include facial areas with limited affective state annotation. Although this kind of setup demonstrated its benefits in online learning, a classroom environment is more complex; thus measuring classroom engagement requires a more comprehensive analysis rather than looking at facial expressions. Sümer et al. collected audiovisual recordings of traditional middle and high school classrooms with a focus on engagement \cite{sumer_multimodal_2021}. Although this dataset considers classroom interaction compared to other datasets, their face recordings are too small to capture facial features, and the dataset is not publicly available.

\section{Five Student-Centric Stages of our Data Collection Process}

\begin{figure}[hb]
    \centering
    \includegraphics[width=\linewidth]{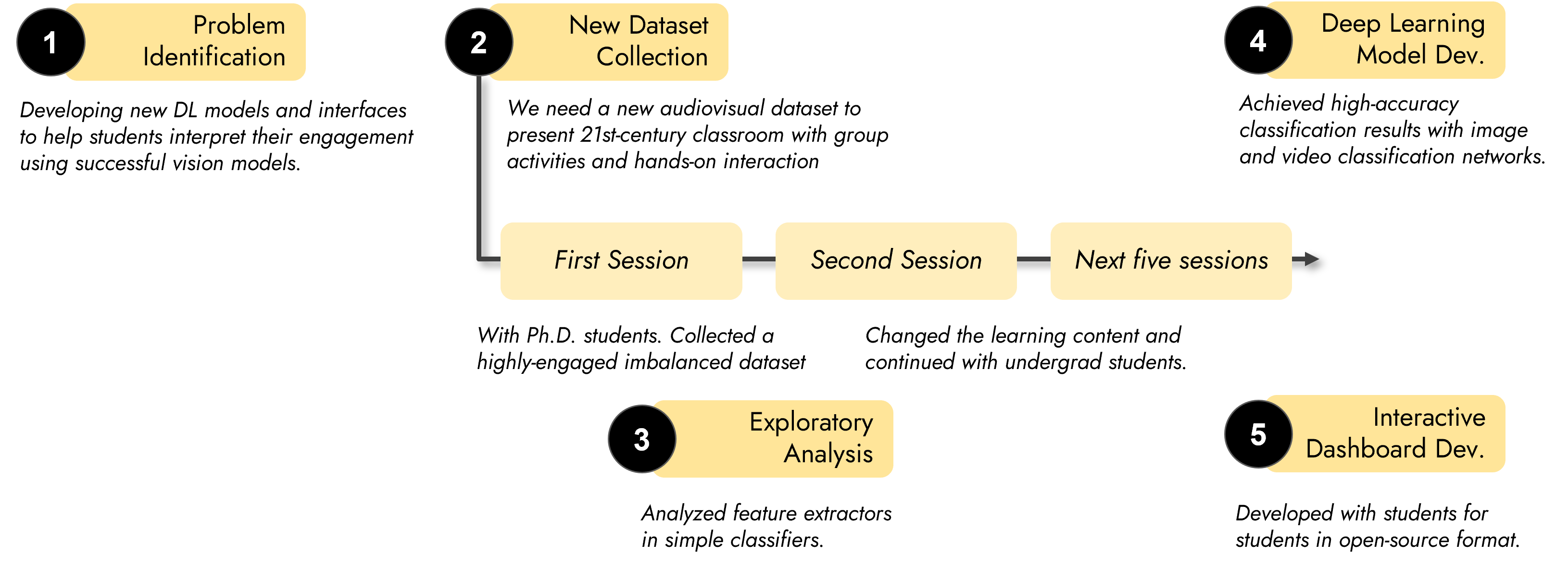}
    \caption{The overall flow of our data collection and model development process. Each package shows the corresponding sections of this paper.}
    \label{fig:overall}
\end{figure}

In our research, we followed five student-centric steps to create our AI-powered engagement prediction system as summarized in Figure \ref{fig:overall}: Data collection, Exploratory Analysis, Model development and fine-tuning, Ethics considerations, and Interactive application for end-result communication.  Each part of the development process consists of unique challenges from a human-centered perspective. This paper presents our development process and open-source materials to accelerate student-centric development in classroom engagement prediction research.

To our knowledge, our end-to-end, video-to-interpretable feature extraction pipeline presents the most comprehensive open-source learning analytics data pipeline that utilizes the most recent state-of-the-art deep learning techniques. Our research outcomes can accelerate the human-centered design of ML-based learning analytics.

\section{Data Collection}
\label{sec:data-collection}

Our dataset consists of group activities of undergraduate and graduate degree students, where they learn new creative coding tools. In the activity design, we aimed to reflect an active classroom setup that follows the 21st-century learning goals. UNESCO defines this 21st-century learning environment as a classroom where students become active and engaged learners, thinkers, and creators \cite{unesco_rethinking_2015}. In this environment, students learn new things, experience hands-on interaction, and reflect on their ideas via constructive discussion. We aimed to curate such a dataset that can be applicable to 21st-century classrooms of K-12 and higher education settings. 

The activity design has slightly changed in an iterative process. But, following a common theme, all groups completed creative coding activities following Youtube tutorials. Then, they also completed some tangible, hands-on activities related to online learning content. Throughout the activities, they also answered some questions that automatically popped up on their tablets to follow their cognitive engagement. They also explored how creative coding applications can be used in research by experiencing some real-life examples.

\begin{figure*}[h]\centering 
\includegraphics[width=.8\linewidth]{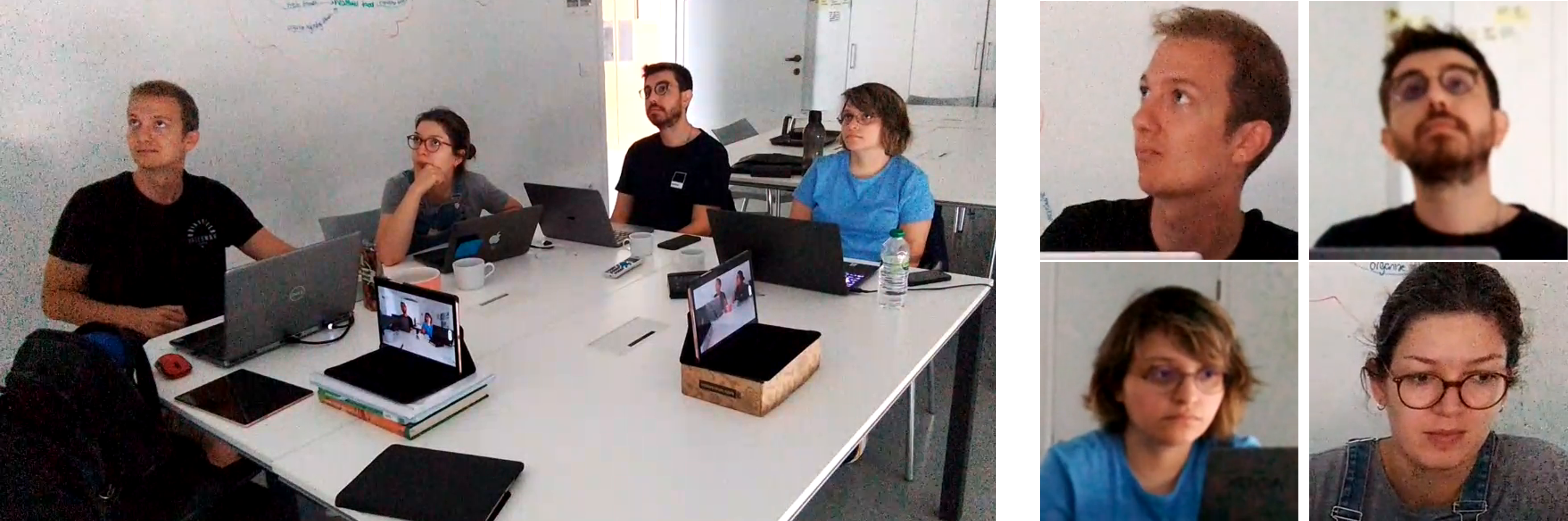}
\caption{(a) One frame from the group-facing camera, (b) One frame from the participant-facing camera.}
\label{fig:frames}
\end{figure*}

\subsection{Participant Selection}

In the participant selection part, we set three requirements to collect life-like data from an active learning environment:

\begin{enumerate}
    \item \textit{Knowing each other before:} We set this requirement to create a similar workflow to a classroom environment. In most classroom learning environments, groups of students already know each other and have things to share in non-related dialog in group activities. 
    \item \textit{Having no or little knowledge on topic:} If a participant already knows the topic, the learning process would be boring and disengaging for this participant.
    \item \textit{Having a teaching experience:} Part of our data analysis includes self-evaluation, which can be a challenging task for the first time. Having previous experience in teaching increased the ability to self-evaluate.
\end{enumerate}

Additionally, collecting data from adult participants allowed us to publish the data publicly. Publishing learning environment data from the K-12 setting requires extra attention, and we could only publish it to the researchers with valid ethical committee permission. As all our participants are older than twenty-two years old, we could publish the dataset in open-source format. The collection complies with GDPR regulations, which means the participant holds every right to their own data, including removing their identity-exposed parts. 

\subsection{Engagement Level Annotation}

Each participant evaluated their self-engagement using a scale between -100 to 100 using the CARMA annotation software \cite{carma_girard2014e}. We prepared an \textit{Engagement Analysis Checklist} to help participants evaluate themselves more accurately. Our analysis checklist is a subset that includes observable inventory items from Wang et al.'s \textit{Classroom Engagement Inventory} \cite{wang_measuring_2014}. Table \ref{table:engagement_checklist} lists the selected checklist items and their engagement category based on Fredricks et al.'s definition. We included fourteen items from this inventory that can be physically observed by an independent evaluator. We reminded participants to regularly control our checklist to remind themselves of the definition of engagement and how classroom engagement is assessed. We also asked participants to complete a survey that asked about their research interests and their motivation to learn the topic.

\newcolumntype{b}{>{\raggedright\arraybackslash}X}
\newcolumntype{s}{>{\raggedright\arraybackslash\hsize=.3\hsize}X}
\begin{table}[h]
\begin{tabularx}{\textwidth}{sb}
 \hline
 Eng. Category & Item \\ [0.5ex] 
 \hline\hline
 Affective & Feeling interested \\ 
 Affective & Feeling proud \\ 
 Affective & Feeling excite \\ 
 Affective & Feeling happy \\ 
 Affective & Feeling amused \\ 
 \hline
 Behavioral & Listening very carefully \\
 Behavioral & Paying attention to the things supposed to remember \\
 Behavioral & Completing the assignment \\
 Behavioral & Getting really involved in class activities \\
 Behavioral & Actively participating in class discussions \\
 Behavioral & Working with other students and learning from each other \\
 \hline
 Cognitive & Trying to figure out where the things went wrong \\
 Cognitive & Searching for information from different places and thinking about how to put it together \\
 Cognitive & Judging the quality of the ideas or work during class activities \\ [1ex] 
 \hline
\end{tabularx}
\caption{Engagement Checklist items and their corresponding categories. We handed out this checklist to our participants while the self-evaluation process to help them memorize the engagement analysis process.}
\label{table:engagement_checklist}
\end{table}

\subsection{Dataset Format and Pre-Processing}

\textbf{Raw videos.} We recorded the videos using three Samsung A8 (2022 Model) tablets in 1280x720 resolution and 30 FPS. One tablet was recording the group from an upper corner. The other two tablets were directly looking at the students that sat side by side. The organization of the cameras can be seen in Figure \ref{fig:frames}. Each video is synchronized by using a clap sound at the beginning. After the clapping moment, we sliced the video based on their respective active materials/topics/discussions. We sliced the videos around ~5-10 minutes to ease the annotation process. Slicing allowed annotators to manage their time efficiently and us to start processing the annotated videos more quickly.

\begin{figure*}[h]
    \centering
    \includegraphics[width=\linewidth]{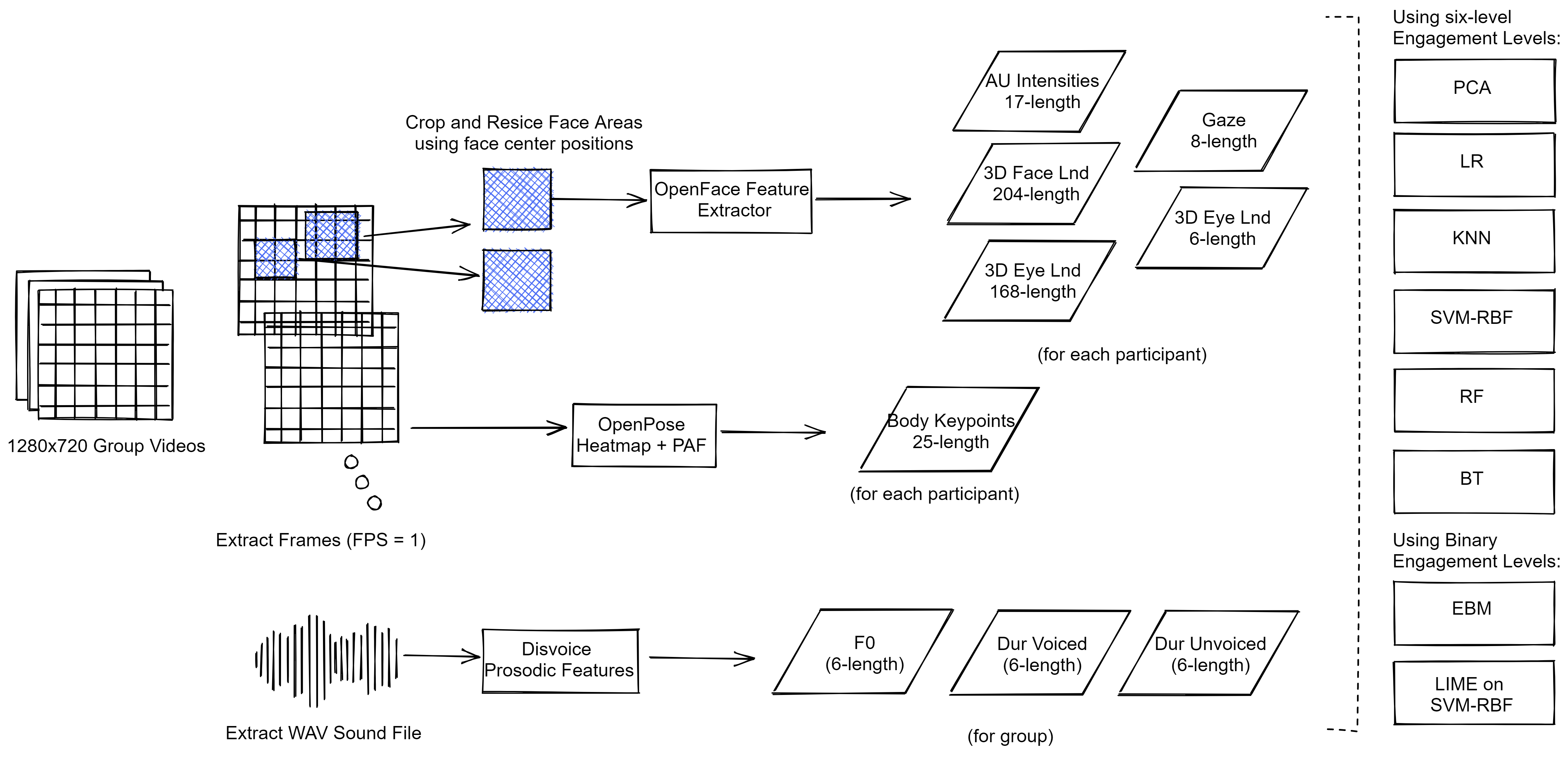}
    \caption{The illustration of our data pre-processing and exploratory analysis pipeline with feature extraction steps.}
    \label{fig:data_pipeline}
\end{figure*}

\textbf{Face Areas.} Then, we cropped and resized the facial areas in order to obtain single-face videos from these two cameras. We used MediaPipe\footnote{https://google.github.io/mediapipe/}'s face detection module to detect the center of the face area. Then, we used FFmpeg\footnote{https://ffmpeg.org/} to crop it by 320px to 320px, which is a commonly used resolution for facial action unit recognition models. In the audio extraction, We applied a noise filter to reduce the background noise during the \textit{wav-conversion} process. In the transcription process, we first run OpenAI’s Whisper model \cite{radford2022robust} to speed up the transcription. Then, we corrected the transcript text. Figure \ref{fig:data_pipeline} summarizes these processing steps.

\textbf{Engagement Level Labels.} Each participant evaluated their self-engagement using a scale between -100 to 100. In addition to individual engagement levels, we determined the overall group engagement levels using different pooling techniques. When a frame has relatively similar scores, such as $[0, 10, 5, 10]$, we applied an average pooling, which results in an average score; for this example, the average is 5. But, if the scores are relatively different, such as $[-60, 0, 60, 100]$, we decided it is a chaotic moment; in other words, it is a disengaging moment. So, we applied a minimum pooling for this frame, which would result in -60 for this example. While determining the similarity, we tested Cappa's quadratic K.

Additionally, we created five engagement levels (highly disengaged, disengaged, moderate, engaged, and highly engaged) for classification tasks. We mapped the values between $[-100, -60)$ as highly disengaged, $[-60, -20)$ as disengaged, $[-20, 20)$ as moderate, $[20, 60)$ as engaged, and $[-60, 100]$ as highly engaged. We used these discrete levels in classification tasks.

\begin{table}[h]
\begin{tabularx}{\textwidth}{sbs}
 \hline
 Feature & Explanation & Link \\ [0.5ex] 
 \hline\hline
 Video Recordings & Three hours of video recordings of first two learning sessions & 
\href{https://www.youtube.com/watch?v=s3JtN4EeS9I\&list=PLrAchnhNnScNtGoxL85y7BSoBj4w_HLIX}{Link} \\ 
 Image Frames &  Image frames of video recordings (FPS = 1). &\href{https://drive.google.com/file/d/1L4USzgfMkTuHK3D2wHGA8QRXQkQQalv2/view?usp=sharing}{Link}\\ 
 Self-Evaluation Scores & The scores are between -100 and 100. Each participant scored their classroom engagement following the \textit{Classroom Engagement Inventory}. &\href{https://github.com/asabuncuoglu13/classroom-engagement-dataset/tree/main/scores/vol02/group/scores}{Link}\\ 
 Engagement Levels & Self-evaluation scores are automatically mapped to five-level classroom engagements: Highly Disengaged, Disengaged, Moderate, Engaged, Highly Engaged.  &\href{https://github.com/asabuncuoglu13/classroom-engagement-dataset/tree/main/scores/vol02/group/levels}{Link} \\
 Video Slices based on Engagement Levels & Slices of videos for the continuous engagement levels.  & 
\href{https://drive.google.com/file/d/1QWhBsqkg4-bahUlsk6UOyqZo6chWZFM5/view?usp=sharing}{Link}\\
 Face Frames & 320x320 image frames of each participant. Centered and cropped. &\href{https://drive.google.com/file/d/1oPEhtvrHOuE2dED7b3UjMgE2CmFxQ-V7/view?usp=sharing}{Link}\\
 OpenFace Features & Combined OpenFace feature CSV for each video. &\href{https://drive.google.com/file/d/14Pe2Ku3vQaEpJvTJOurvjW07ppZmDF13/view?usp=sharing}{Link} \\
 Body Keypoints & Combined body keypoint positions for all frames. &\href{https://drive.google.com/file/d/1e623GY1j90XfRZPmDybzkDMhdmAgawym/view?usp=sharing}{Link} \\
 OpenPose Features & OpenPose Heatmap for different scales and keypoint JSON for each frame. &\href{https://drive.google.com/file/d/1_4PlZ4fBKJUvMst9FGMnIgpLPaXTRbOx/view?usp=sharing}{Link} \\
 Audio & WAV file - We applied noise reduction on the original audio.  &\href{https://drive.google.com/drive/folders/1Qw9c1ci2a2hlD8l1gRdTknJKHsAPedqt?usp=sharing}{Link} \\
 Transcript & Automatically transcripted using OpenAI's Whisper, then manually checked by humans. &\href{https://drive.google.com/drive/folders/1ojrDirqw2HfpbjDPsXQN5u63SlZXerYJ?usp=sharing}{Link} \\[1ex] 
 \hline
\end{tabularx}
\caption{Media and feature set for each session of the dataset. This table contains links for an example session which we used in the exploratory analysis (Session 2). All dataset and features can be downloaded via our \href{}{Download Script}}
\label{table:statistics}
\end{table}

\subsection{Exploratory Analysis of First Batch}

Our analysis also progressed in an iterative fashion. After the first two data collection studies, we conducted an early analysis to update the data collection process and resolve the issues as soon as possible. The main issue of the first batch was the imbalance of the engagement levels. The ~70\% of the frames were labelled as highly engaged in self evaluation, which presented a highly imbalanced dataaset. Following the exploratory analysis of the first two sessions, we updated our learning setup and activity flow to reduce this imbalance. The new activity flow achieved more balanced data that shows a normal-distribution like engagement-level category distribution. In this section, we presented our analysis for the first session i which three participants are anonymized as B, C and Y initials.

\subsubsection{Methodology}

We used OpenFace \cite{baltrusaitis_cross-dataset_2015} to extract facial action units and OpenPose \cite{cao_openpose_2019} to obtain the joint positions of students. OpenFace's Face Feature Extractor yields 1562 features per vector. We analyzed five continuous feature sets: AU Intensities, 3D Eye Landmarks, 3D Face Landmarks, Gaze Directions, and Head Pose, and one discrete AU Classes Feature Set. We extracted a feature vector for each frame (FPS = 1). The resulting features are interpretable in terms of learning analytics, as the Feature Extractor results in features like eye gaze, head position, action units, etc. Before feeding the feature set to machine learning models, we cleaned the data by thresholding the confidence score of face features. We removed all feature vectors that have less than a $0.5$ confidence score. We also removed the vector if the row contains only zeros. Out of 6969 features, we trained our models with 5351 feature vectors. Using these features, we conducted a three-step analysis strategy in this exploratory analysis: 

\begin{enumerate}
    \item For each feature set, we applied PCA (Principal Component Analysis) to see the possible linearly separable features that can increase the overall engagement level classification. Principal Component Analysis (PCA) aims to reduce the dimensionality of a set of data consisting of variables correlated with each other while keeping the variation in the dataset \cite{Jolliffe2016}. The variation in the principal components decreases while moving from the first component to the last one, and we can measure the variation explained with the PCA's dimensionality reduction by cumulative explained variation. As Open Face's Action Units result in discrete observations, we utilized MCA (Multi Correspondance Analysis) instead of PCA.
    \item We trained five classifiers, Logistic Regression (LR), k-Nearest Neighbours (kNN), Radial-Basis-Function Kernel Support Vector Machine (RBF-SVM),  Random Forest (RF), and Boosted Trees (BT) using individual and combined feature sets to analyze the impact of features and model characteristics on engagement level prediction. 
    \item Finally, we conducted explainability experiments to find the impact of individual features and test the combination of these important features on the same classifiers. We utilized InterpretML \cite{interpretml}, an open-source Python library, to explain the behavior of existing systems using LIME. A drawback of the current implementation of InterpretML is its limited availability for multi-class classification. So, we produced interpretable rules on binary predictions. We run these algorithms on a binarized version of the data using the classifier. 
\end{enumerate}

Our open-source repository contains codes for conducting these exploratory analysis step by step. Researchers can easily conduct analysis on their custom dataset or customize our code to conduct different experiments on our dataset.

\subsubsection{Dimensionality Reduction}

\begin{figure*}[h]
    \centering
    \includegraphics[width=\linewidth]{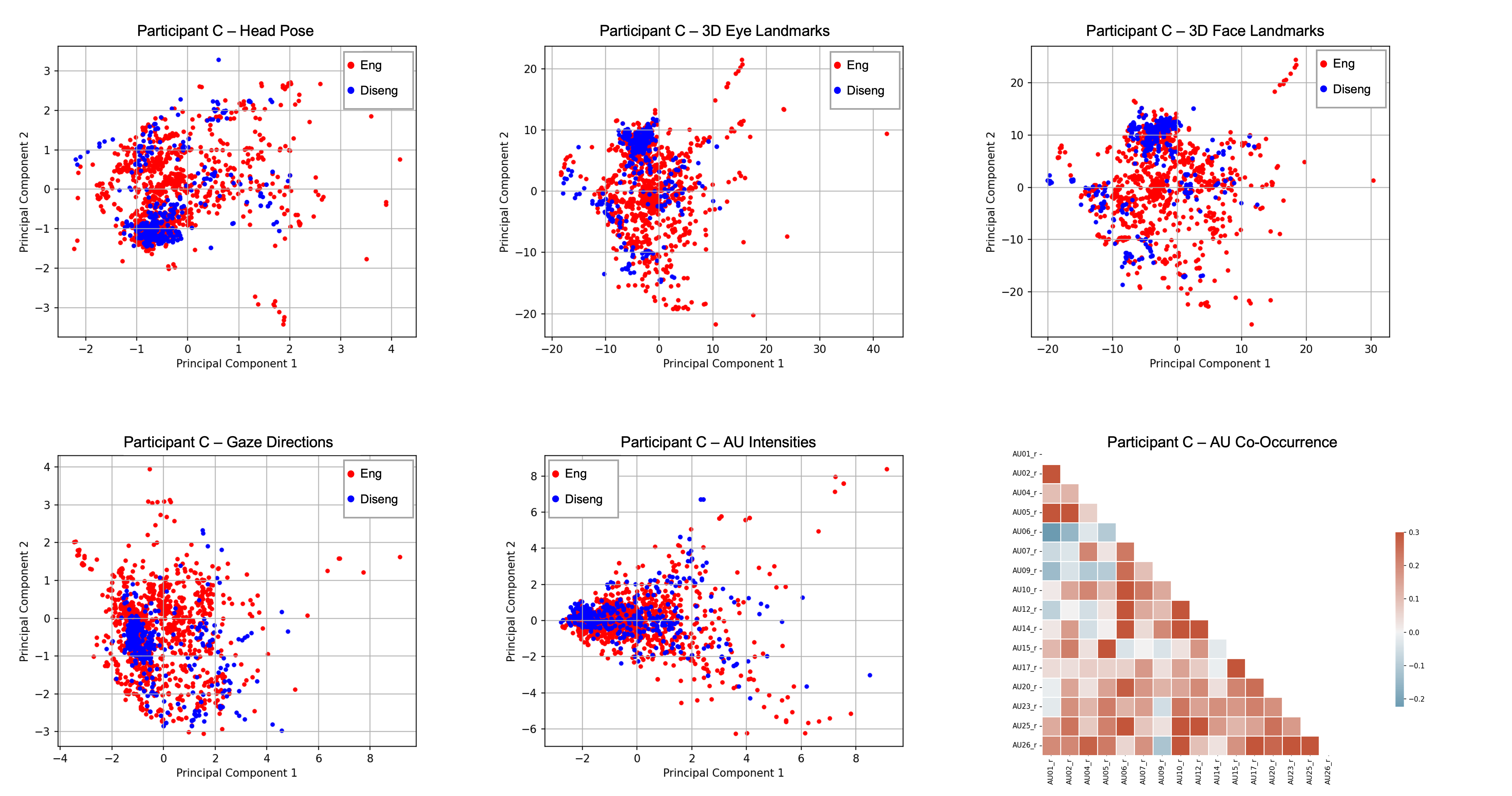}
    \caption{Example figures from 2-component PCA analysis of Participant C. Best for electronic distribution. }
    \label{fig:pca}
\end{figure*}

Figure \ref{fig:pca} demonstrates an example PCA analysis for Participant C's OpenFace features. As the figures clearly show, the feature space is not linearly separable, but Head Pose, 3D Eye, and Face Landmarks formed some clusters compared to other features. We also presented the AU co-occurrence of Participant C in this figure. The highest co-occurrence value for all participants is 0.3, which shows no clear indication of correlation between action units for this experiment.

In addition to PCA, t-distributed stochastic neighbor embedding (t-SNE) and Uniform Manifold Approximation and Projection (UMAP) are non-linear dimensionality reduction techniques that focus on preserving the local structure of the data \cite{tsne, umap}. They are popular methods for visualizing high-dimensional space, but their non-linear nature can result in misleading output. We believe an effective way to explore this high-dimensionality space is to allow readers to explore the data with their custom parameters, which they can find a balance between local and global structure. So, we published the feature sets in \textit{Embedding Projector}-friendly format (You can play with the format using the sample features from Session 2:
\href{https://drive.google.com/file/d/16d-BUGooIwTLy-c1Og7Ggdu0GfZZgzYy/view?usp=sharing}{Embeddings - Google Drive}). Tensorflow’s Embedding Projector \cite{embedding-projector} is a web application to visualize high-dimensional format using PCA, t-SNE, or UMAP algorithms interactively. Through the Embedding projector, researchers can play with the dimensionality-reduction parameters.

\subsubsection{Classification using Individual and Combined Features}

In the classifier performance evaluation, we calculated the weighted F1-score metric with our 80-20 split dataset. The F1 score is a harmonic mean of precision and recall. The weighted F1-score calculates scores for each label and finds their weighted average, which provides a more informed decision when data is imbalanced.

\textbf{Classification using OpenFace Features:} Table \ref{table:openface-classifiers} reports the weighted F1 scores for five OpenFace features and selected combined features that we hypothesized combination could yield better accuracy: 3D Eye Landmark, 3D Face Landmark, Gaze Directions, Head Pose, and AU Intensities.

\begin{longtable}{|l|l|l|l|l|l|l|}
\hline
                        &                 & LR     & KNN   & RBF-SVM & RF    & BT    \\ \hline
\endfirsthead
\endhead
\multirow{6}{*}{B}   & 3D Eye Lndm     & 0.418  & 0.576 & 0.339   & 0.301 & 0.410  \\ \cline{2-7} 
                        & 3D Face Lndm    & 0.365  & 0.577 & 0.391   & 0.316 & 0.479 \\ \cline{2-7} 
                        & Gaze Directions & 0.294  & 0.384 & 0.385   & 0.310  & 0.384 \\ \cline{2-7} 
                        & Head Pose       & 0.287  & 0.555 & 0.585   & 0.316 & 0.507 \\ \cline{2-7} 
                        & AU Intensities  & 0.331  & 0.556 & 0.394   & 0.296 & 0.479 \\ \cline{2-7} 
                        & All-OpenFace    & 0.418  & 0.576 & 0.338   & 0.306 & 0.501 \\ \hline
\multirow{6}{*}{C}  & 3D Eye Lndm     & 0.399 & 0.596 & 0.515   & 0.440  & 0.526 \\ \cline{2-7} 
                        & 3D Face Lndm    & 0.468  & 0.686 & 0.564   & 0.448 & 0.597 \\ \cline{2-7} 
                        & Gaze Directions & 0.294  & 0.496 & 0.467   & 0.379 & 0.453 \\ \cline{2-7} 
                        & Head Pose       & 0.370   & 0.688 & 0.659   & 0.452 & 0.584 \\ \cline{2-7} 
                        & AU Intensities  & 0.318  & 0.609 & 0.479   & 0.318 & 0.511 \\ \cline{2-7} 
                        & All-OpenFace    & 0.516  & 0.684 & 0.354   & 0.445 & 0.606 \\ \hline
\multirow{6}{*}{Y} & 3D Eye Lndm     & 0.472  & 0.558 & 0.390    & 0.488 & 0.525 \\ \cline{2-7} 
                        & 3D Face Lndm    & 0.499  & 0.662 & 0.446   & 0.504 & 0.591 \\ \cline{2-7} 
                        & Gaze Directions & 0.380   & 0.440  & 0.459   & 0.414 & 0.448 \\ \cline{2-7}
                        & Head Pose       & 0.411  & 0.648 & 0.631   & 0.518 & 0.524 \\ \cline{2-7} 
                        & AU Intensities  & 0.390  & 0.589 & 0.518   & 0.432 & 0.517 \\ \cline{2-7} 
                        & All-OpenFace    & 0.517  & 0.645 & 0.270    & 0.508 & 0.597 \\ \hline
\caption{F1 Scores of classifiers on individual OpenFace features (3D Eye and Face Landmarks, Gaze Directions, Head Pose, AU Intensities) and combination of all features for each participant. B, C, and Y are different participants.  }
\label{table:openface-classifiers}\\
\end{longtable}

\textbf{Classification using OpenPose:} Table \ref{table:openpose-classifiers} reports the weighted F1 scores for each participant's OpenPose joint keypoints. OpenPose returns twenty-five keypoint positions for each person in three axes.

\begin{longtable}{|l|l|l|l|l|l|}
\hline
             & LR    & KNN   & RBF-SVM & RF    & BT    \\ \hline
\endfirsthead
\endhead
P1           & 0.363 & 0.492 & 0.370   & 0.267 & 0.510 \\ \hline
P2           & 0.332 & 0.461 & 0.363   & 0.264 & 0.459 \\ \hline
P3           & 0.349 & 0.463 & 0.388   & 0.297 & 0.465 \\ \hline
P4           & 0.367 & 0.517 & 0.461   & 0.281 & 0.524 \\ \hline
All Features & 0.474 & 0.439 & 0.324   & 0.256 & 0.517 \\ \hline
\caption{F1 Scores of OpenPose keypoints for each participant.}
\label{table:openpose-classifiers}\\
\end{longtable}

\subsection{InterpretML Results}

We run EBM and LIME on SVM-RBF for five feature sets for each participant similar to the classifier experiments. LIME and EBM can only run on binarized datasets, so we created an additional engaged-disengaged labels for binary classification task. In these experiments, we observed that action unit scores and head position information have the highest impact on the model at the individual level. An overall sample analysis of explainability scores looks like the rules in Table \ref{table:exp}.

\begin{table}[hb]
  \begin{tabular}{ |p{.3\linewidth}|p{.7\linewidth}|}
    \hline
    \textbf{Component} & \textbf{Rule} \\ 
    \hline
    Data Stats & The binarized engagement level of Session 2 have 4977 \textit{Engaged} observations and 944 \textit{Disengaged} observations. \\ \hline
    Highest Score & The highest F1 score with the binarized dataset occurred when we fed 3D Face and Head Pose information, which yielded 0.84.  \\ \hline
    Action Units & Combination of action units had more impact than individual futures. Combinations with AU5 (upper lid raiser) had more impact than other action units. \\ \hline
    Pose Features & All other features could show impact when only combined with other features. The individual features could not show a significant impact ($<= 0.1$). In all classifiers lower-body parts (RSmallToe, LSmallToe, RHell, etc.) had the highest impact on decision-making. \\ \hline
    Morris Sensitivity \cite{noauthor_introduction_2007} & Similarly, Morris sensitivity values only yielded meaningful values for gaze direction and head pose features. In the head pose, we also observed the impact of y-axis values. The sensitivity values for pose features did not yield meaningful values (could not converge), which indicates the individual impact of features is very low. \\ \hline
\end{tabular}
\caption{A sample analysis of LIME feature values. A different model can yield different important components. So, for different models at different inference times these components might change.}
\label{table:exp}
\end{table}

Researchers can access our full interpretability reports and generate their interpretability reports on our Github Repo.

\subsection{Final Statistics} 

After completing the exploratory analysis and updated the activity flow, we continued our user studies with additional five sessions. The released dataset contains 26540 frames (FPS = 1, ~7.5 hours) of audiovisual recordings with self-evaluation scores for each frame. For each frame, individual faces are centered and cropped to 320x320 images. For each frame, with two-second paddings, we also sliced ten-second-long video clips to feed video action recognition networks. We automatically extracted dialog transcripts using Whisper \cite{whisper} and manually checked these transcripts, thanks to volunteers. Table \ref{table:statistics} shows the explanations and open-source links for available features of this dataset.

We released all the parts of the datasets where the participants allowed the use of their visuals and audio. Four out of thirty-four students did not allow sharing the video recordings, so we cannot publicly share three of the seven videos. However, researchers holding \textit{Ethics Committee Approval} can request the private links. In the end, the frame counts for each engagement level in a five-level classification task are:

\begin{itemize}
    \item \textbf{Highly engaged:} 5914 (22\% of all data)
    \item \textbf{Engaged:} 5489 (20\% of all data
    \item \textbf{Moderate:} 7888 (30\% of all data)
    \item \textbf{Disengaged:} 4271 (16\% of all data)
    \item \textbf{Highly Disengaged:} 3084 (12\% of all data)
\end{itemize}

At the end of the data collection process, we started developing baseline architectures that use the image frames and video clips.

\section{Baseline Architectures for Engagement Prediction}
\label{sec:dl-models}

Following the data collection and pre-processing steps, we developed baseline deep learning architectures to classify engagement levels. Previous research demonstrated that using mid-level features using pre-trained networks is an effective method in video understanding \cite{acar_comprehensive_2017}. Following this insight, we tested different mid-level representations and feature sets in our baseline architectures. In these experiments, we tested sequential learning architectures (GRUs and LSTMs) architectures with different feature representations (VGGFace, MobileNet, MoViNet).  In this paper, we presented classification architectures with two base data feeding models. The first set of architectures takes 320x320px size image inputs that contain cropped and centered face areas with individual engagement levels as labels. The second set of architectures takes 1280x720px size 10-second video clip inputs with aggregate group engagement levels as labels.

For all experiments, we split the data into 80\% training (20\% of it is validation) - 20\% test data.

\subsection{Models for Centered Face Input}

These models make predictions using 320x320px images, which contain cropped face areas in the center of the image. We present two baseline architectures that test different pre-trained weight initialization and activation functions: (1) The first architecture tests two pre-trained weight initializations, VGG19-Face and MobileNet. (2) The second architecture uses the MobileNet-based model with MaxOut \cite{wu_light_2018} and ReLU activation functions.

\subsubsection{Experiment 1: VGG19-Face vs. MobileNet for pre-trained weight initialization}

On top of pre-trained networks, we trained a densely connected NN with ReLU activation. The network uses sparse categorical cross entropy with Adam optimizer (lr = 0.0001). We trained the network for 50 epochs. Then, we also tested fine-tuning the pre-trained weights to further improve the model performance. In the fine-tuning step, we used the RMSProp optimizer to restrict the oscillations that might occur in the gradient descent. We trained the network with the fine-tuning step with another 50 epochs. Using MobileNet as pre-trained weight initialization demonstrated better performance and accuracy for each participant’s engagement classification. Thus, we tested the second model with only MobileNet weights.

\subsubsection{Experiment 2: MaxOut Dense Layer vs. ReLU Dense Layer for Prediction Head}

As we elaborated in Section \ref{sec:data-collection}, the self-evaluation labels can be unreliable and noisy. So, we also included MaxOut activation that showed promising results for noisy labels in previous face recognition tasks \cite{wu_light_2018}. For the MobileNet model, we tested ReLu activation and MaxOut activation for the prediction heads with a similar implementation to the LightCNN network. The MaxOut implementation, as expected, demonstrated better accuracy. 

\subsubsection{Results}

The VGG19-Face is specifically trained to recognize faces, so we expected to observe a better weight initialization when we started our model training with VGG19-Face. However, MobileNet demonstrated significantly better performance in validation and test accuracies. So, we can deduce that MobileNet is a better feature extractor compared to VGG19-Face, even when the image area mainly contains faces. Thus, we continued with MobileNet as a feature extractor while experimenting with activation layers. In this experiment, MaxOut implementation demonstrated better accuracy as it reduced the possibility of overfitting and showed better accuracy in unreliable data. Table \ref{tab:face-area-model-acc} reports the validation accuracy before and after fine-tuning and test accuracy of the final model for the MobileNet feature extractor with MaxOut activation, which performed the best accuracies among all experiments for the centered-face input.

\begin{longtable}{|l|l|l|l|l|}
\hline
Session & Person & Val-Acc & Val-Acc-Fine-Tune & Test-Acc \\ \hline
\endfirsthead
\endhead
4 & 1 & 0.39 & 0.43 & 0.50 \\ \hline
4 & 2 & 0.60 & 0.62 & 0.56 \\ \hline
4 & 3 & 0.54 & 0.53 & 0.41 \\ \hline
4 & 4 & 0.60 & 0.78 & 0.91 \\ \hline
4 & 5 & 0.60 & 0.66 & 0.72 \\ \hline
5 & 1 & 0.51 & 0.71 & 0.69 \\ \hline
5 & 2 & 0.60 & 0.74 & 0.81 \\ \hline
5 & 3 & 0.44 & 0.49 & 0.50 \\ \hline
5 & 4 & 0.48 & 0.57 & 0.53 \\ \hline
5 & 5 & 0.58 & 0.78 & 0.59 \\ \hline
6 & 1 & 0.38 & 0.46 & 0.47 \\ \hline
6 & 2 & 0.39 & 0.54 & 0.41 \\ \hline
6 & 3 & 0.57 & 0.47 & 0.47 \\ \hline
6 & 4 & 0.66 & 0.72 & 0.78 \\ \hline
6 & 5 & 0.83 & 0.82 & 0.84 \\ \hline
7 & 1 & 0.64 & 0.73 & 0.66 \\ \hline
7 & 2 & 0.48 & 0.53 & 0.63 \\ \hline
7 & 3 & 0.72 & 0.78 & 0.78 \\ \hline
7 & 4 & 0.47 & 0.77 & 0.72 \\ \hline
7 & 5 & 0.69 & 0.85 & 0.78 \\ \hline
8 & 1 & 0.43 & 0.44 & 0.41 \\ \hline
8 & 2 & 0.37 & 0.49 & 0.60 \\ \hline
8 & 3 & 0.45 & 0.45 & 0.59 \\ \hline
8 & 4 & 0.46 & 0.49 & 0.53 \\ \hline
8 & 5 & 0.43 & 0.46 & 0.53 \\ \hline
\caption{Validation and test accuracies of the MobileNet feature extractor with MaxOut activation for each participant.}
\label{tab:face-area-model-acc}\\
\end{longtable}

\subsection{Models with Video Input}

These models take 10-second long video clips that have 1280x720 resolution. In the data preparation step, we had different options to experiment with. For example, one option was to slice videos based on consecutive engagement levels. Following this option, we obtained 1521 video slices. The shortest video clip was one second, and the longest video clip was 146 seconds. This method of slicing videos did not yield a balanced dataset in terms of quantity, label distribution, and duration. So, we prepared 10-second slices with 2-second paddings. Intuitively, we can say that if we select a random frame label, the 10-second interval (5 seconds before and 5 seconds after) has the same or close label. So, we created 10-second intervals that carries labels of their center frames. This way, we obtained 25K video clips that have 2 seconds of padding.

\begin{figure}[hb]
    \centering
    \includegraphics[width=.9\linewidth]{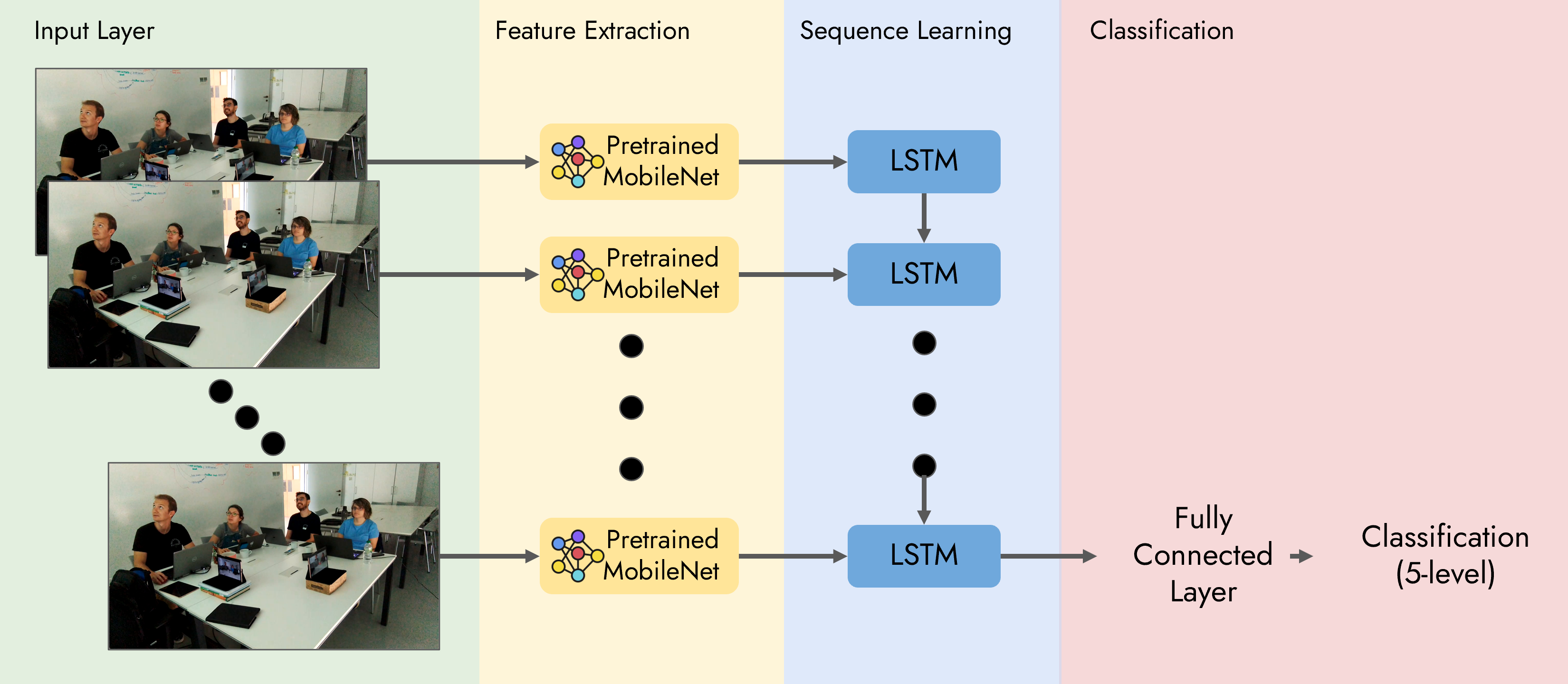}
    \caption{An overview of sequence-based learning using MobileNet features of individual frames.}
    \label{fig:face-model}
\end{figure}

Using this data, we trained two different architectures with different base feature extractors: (1) The first experiment tests LSTM-based models with MobileNet and VGG19Face feature extractors. (2) The second experiment utilizes the MoViNet model as the base video prediction model and applies fine-tuning to learn from few videos.

\subsubsection{Experiment 1: RNN with CNN Feature Extractors}

Our first model uses CNN-based models (VGG19Face and MobileNet) that we previously utilized in face-area-feeding networks to LSTM-based network as seen in Figure \ref{fig:face-model}. The experiment run for twenty epochs with the batch size 60 (60 video clips). The network uses sparse categorical cross entropy with Adam optimizer (lr = 0.0001).

\subsubsection{Experiment 2: MoviNet A0 and A4}

Although CNN-RNN models have been dominantly used in video prediction models, a 2D-frame-based classifier lacks representation of temporal context. Tran et al. demonstrated that 3D convolutions have better performance in capturing spatiotemporal information compared to previous models \cite{tran_3dconv_2017}. In the second model, we used a descendant of a 3D convolution-based architecture, MoViNet.  We used MoViNet A0 and A4 base models, which showed promising results on the Kinetics-600 classification task \cite{kondratyuk_movinets_2021}. On top of MoViNet models, we trained a GRU-based classification head.  The experiment run for twenty epochs with the batch size 20 (20 video clips). The network uses sparse categorical cross entropy with Adam optimizer (lr = 0.0001).

\begin{figure}
    \centering
    \includegraphics[width=.6\linewidth]{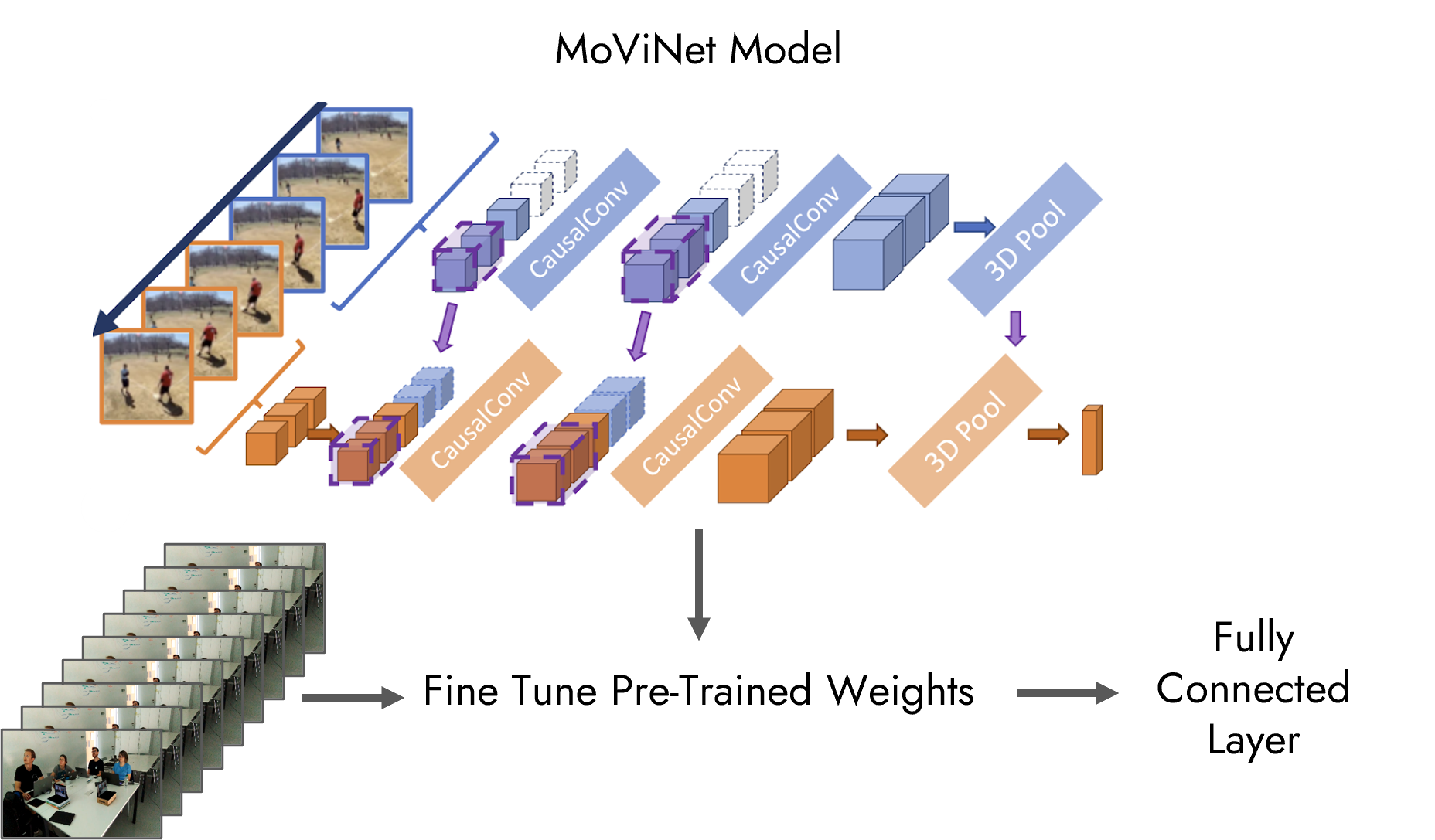}
    \caption{The high-level presentation of MoViNet-based classification using fine-tuning of MoViNet models. MoviNet architecture is obtained from \cite{kondratyuk_movinets_2021}.}
    \label{fig:video-model}
\end{figure}

\subsubsection{Results}

CNN-LSTM architecture yielded poor accuracy, compared to MoViNet-based models. Table \ref{table:video-results} shows the validation and test accuracy results of these architectures. In the CNN-LSTM architecture VGG19Face and MobiNet showed similar performance in feature representation. Overall, the 2D-frame based representation could not yield any meaningful results. However, fine-tuning MoViNets yielded promising results. MoViNet-A4 base classifier could achive 0.68 accuracy on test set.

\begin{table}[hb]
  \begin{tabular}{ |p{.5\linewidth}|p{.2\linewidth}|p{.2\linewidth}|}
    \hline
    \textbf{Architecture} & \textbf{Val-Acc} & \textbf{Test-Acc} \\  \hline
    CNN-LSTM (MobileNet) & 0.46 & 0.41 \\ \hline
    CNN-LSTM (VGG) & 0.48 & 0.41 \\ \hline
    MoViNet (A0) & 0.71 & 0.63 \\ \hline
    MoViNet (A4) & 0.76 & 0.68 \\ \hline
     
\end{tabular}
\caption{Different video action classification model architectures and their validation and test accuracies.}
\label{table:video-results}
\end{table}

%----------------------------------------------------------
% DISCUSSION
%----------------------------------------------------------
\section{Discussion}
\label{sec:discussion}

From data collection to dashboard development, we followed a human-centered approach in all steps of our machine learning pipeline for predicting student engagement levels. In this section, we hold a discussion over the main findings from our data collection, analysis, and model development steps with a student-centric view.

\subsection{Common Patterns in Engagement Levels}

\textbf{Engaging patterns:} Our activity flow consisted of two parts where participants first learned to use creative coding applications following Youtube tutorials, then they completed a tangible activity that they followed using an online tutorial. While trying to complete the hands-on task, the participants used tangible materials and a tablet computer. During this interaction, the disengagement/engagement ratio of this part becomes low compared to the video-watching activity. This increase in engagement also resonates with the previous research in tangible interaction research. We promote adding more learning activity types, such as peer-to-peer learning and flipped classrooms, to increase the variety in the dataset and help educators interpret the patterns in different learning tasks.

\textbf{Disengaging patterns:} We can list two main patterns for disengagement: (1) When the instructor in the video started to talk about future work, participants tended to decrease their engagement scores. This situation might also occur since the timing of these conversations generally happens at the end of the lecture time. (2) When participants faced a technical failure, they did not feel disengaged but set their engagement to a moderate level. Listing these repeating patterns in engagement annotation can help educators to plan their lectures accordingly.

\textbf{Outliers:} Through the evaluations, we also faced unpredictable annotations. For example, when students were laughing and enjoying, one can expect to label it as an engaging moment. However, our engagement checklist also considers cognitive and behavioral engagement; thus, these enjoyable moments should relate to the course topic. However, we know that these moments help students keep their engagement level long-term. So, labeling and recognizing these moments are challenging both for ML models and teachers. Most of the false positive examples in our dataset consist of these moments.

\subsection{Analyzing Unreliability in Self-Evaluation Process}

Through the learning activities in the data collection process, we asked participants questions about the video content to check if their engagement levels and correct answers would match. The main goal of asking these questions was to see if their self-evaluation engagement score was in parallel with their answers. We aimed to achieve a reliability check with this kind of mechanism. After collecting the engagement level annotations, we checked the engagement levels one minute before and after the timestamp of this question. 

The first question was, “What is p5.js?”. Two participants answered incorrectly. These students also labeled their engagement scores as negative. We also observed that our group activity model could also predict their engagement level correctly. But the second question was tricky. We asked, “Which platform can you use to code p5.js?” Only five of them answered correctly. They selected the platform that they used in the activities. But, the correct answer required choosing multiple platforms. In the evaluation part, most of them labeled themselves as ‘engaged,’ which indicates that most of our participants tend to label their engagement with their overall behavior rather than giving careful attention to the lecture.

Although this test can only be used for an exploratory and limited experiment for reliability, future researchers can follow a similar process to understand the reliability of their data collection process.

\subsection{Interpreting Observable Features of First Batch}

Our early experiments involving OpenFace and OpenPose features revealed that 3D Face, Eye Landmarks, and HeadPose features significantly improved accuracy compared to other facial, pose, and voice features. Explainability experiments also demonstrated that the combination of these features had the highest F1 score compared to other individual and combined features. Although gaze directions can reveal insights about shared attention between the students, the least successful models were trained with gaze directions. Previous research demonstrated that mutually shared attention could reveal some insights for classroom engagements. Still, we observed that gaze direction could also falsely guide the models when students are just in a mind-wandering state. 

Through the analysis, we also check if our data features resonate with the literature. We expected to achieve the best results by either using AU intensities or combining AU intensities with other features, as the previous research could effectively predict affective states using action units. For example, the activation of AU4, AU7 and AU12 can indicate “Boredom”, and AU1, AU4, AU7 and AU12 can reveal “Confusion.” \cite{dewan_engagement_2019}.  But, when we checked if the students annotated "disengaged" when the action units were active for the "boredom" state, we could not achieve any significant result. To check the OpenFace's reliability in predicting action units, we also conducted a random frame check where we picked random frames and manually checked if OpenFace produced sound AU intensity values. In this examination, we concluded that OpenFace produces reasonable AU intensities, but we realized that predicting an engagement score using a single observation is a challenging task. For example, while they were eating some snacks while watching the video or looking serious while trying to solve a task on a computer, AU intensities were producing the same scores when they were disengaged. 

Head pose, to our surprise, had a significant impact on all classification experiments. When we analyzed the feature importance reports, we observed that y-axis values are the main components. While watching the videos, we also observed that our participants (especially participant B) tended to node while listening carefully. We can also generalize this outcome intuitively and say that nodding and similar head pose behaviors can be key components of engagement analysis.

Before our analysis, we expected to see the major contribution of OpenPose features would be the change in the hand movements, head positions, and the distance between the participants. At the beginning of the analysis, we were planning to remove the lower body keypoints, as most of the points are not seen but predicted by the model. But, we decided to keep it, as participants stand up, talk to each other, or leave the class, which shows the change in the engagement levels. In the explainability analysis, we see that these keypoints had a major impact on EBM's decisions. So, future work should include either better keypoint prediction of occluded leg movements or increase the camera field of view to see the whole body.

\subsection{Effective Data Analysis Practices} 

\textbf{Combining features based on feature importance:}  Combining features from different knowledge sources can enhance classification accuracy significantly, but it can also cause overfitting. In the individual classification tasks, we observed that \textbf{3D Face Landmarks} and \textbf{Head Position} information are the main indicators of successful engagement level classification tasks, and combining them resulted in the highest scores for both classification and explainability experiments. Although combining features can improve the classifier performance significantly, simply adding all features together does not make the expected impact, as the classification algorithms struggle to find a converging path. So, while combining the features, we need to select relatively small-sized features that show high orthogonality with each other.

\textbf{Crafting new features:} We also tested creating additional features that might also be relevant to real-life situations. For example, we hypothesized that when students become closer in a classroom, it could be an indicator of engagement. So, we calculated the Euclidean distance between the participants' body keypoints and trained the classifiers with this closeness feature. But, the experiments did not yield the expected classification accuracy.

\textbf{Adding more and more features:}  Increasing the future count with a limited dataset size results in weak accuracy scores, as converging to an optimal solution is a challenge. Selecting a subset of features is much more useful than feeding all networks to reduce the training and inference time. In our experiments, a 400-length feature vector took approximately spend 10-times more training time than a 200-length feature vector.

\subsection{Testing Deep Learning Architectures}

Our experiments demonstrated that carefully selecting pre-trained feature extractors in a transfer learning approach is the most important factor in achieving good accuracy. Currently, the most effective and high-accuracy models are run on top of MobileNet (for face-area-fed models) and MoViNet (group-activity-video-fed models) architectures. We encourage researchers to test different feature extractors and mid-level representations that could increase model accuracy significantly.

In addition, the dataset is limited in terms of generalizing the capabilities of the models. As we elaborated in more detail in Section \ref{subsec:limitations}, the dataset is collected in a university setting where participants are from the same department. This limits the affective, cultural, behavioral, cognitive, and pedagogical representation of the model. So, the adopters and researchers should investigate the ethical considerations related to representability and bias.

\subsection{Ethical Considerations}

At the beginning of our development process, our vision for the classroom scenario was supporting students and teachers in observing engagement patterns and interpreting their engagement levels. Yet, one can be concerned about using the overall system as a classroom surveillance tool, which can process personally-identifiable data that classifies behavior, attitudes, and preferences. Privacy concerns has been a prominent challenge in the learning analytics field \cite{lang_quantifying_2020, slade_learning_2019, arnold_student_2017, prinsloo_elephant_2017, drachsler_privacy_2016, prinsloo_student_2015}. Williamson et al. list four emerging challenges while developing LADs \cite{williamson_review_2022}: (1) Protecting participants' privacy while also including enough demographic information, (2) Surveillance concerns, (3) Neglecting pedagogies that fall outside of the dominant narrative and (4) Making LADs maintainable in terms of software development. For each of these emerging issues, we summarize our approach to help researchers and practitioners utilizing our prediction pipeline.

\begin{enumerate}
    \item \textbf{Protecting participant identities:} Our pipeline does not require collecting any demographic information. If a third party adapts our pipeline, they can run the models anonymously without requiring any identifiers. Throughout the use of our system, teachers have access to all data, but students do not have access to other students' engagement levels. Currently, we identified two main users: Teachers and Students. Other stakeholders such as policy-makers might need demographic information to make nationwide decisions. At this point, the engagement data should be aggregated in a privacy-preserving way to protect personal identifiers \cite{applebaum_collaborative_2010}. 
    \item \textbf{Adressing surveillance concerns:} Our system only give access teachers to personally identifiable data of their students. A student can only see video data from other members of the study group. Our deep learning architectures do not submit any information to third-party software. For each step of our system, we included a step-by-step explanation of data usage in an end-user-readable way, and we also suggested this approach to adapters of our system.
    \item \textbf{Considering implicit pedagogies:} By using our overall system, students and teachers can explore their learning patterns that fall behind the dominant narrative. The final dashboard aims to help students interpret their engagement levels. If other researchers and adapters of the system intend to give suggestions based on their pedagogic approach, they should carefully support their arguments by showing direct links to ML model features.
    \item \textbf{Maintaining the software:} Making our dataset, pipeline, models and dashboard open-source and presenting our dashboard on Observable improves the software maintainability significantly. Using Observable, researchers and developers can fork our interactive notebook and create custom dashboards based on their needs. They can also access our data pre-processing scripts and DL model training codes through the project's GitHub repo (\url{https://github.com/asabuncuoglu13/classroom-engagement-dataset}), which is currently active and open-source.
\end{enumerate} 

Lastly, we shared these resources with a share-a-like license, so adapters should also make their code open-source to increase maintainability and sustain the ethical considerations.

\subsection{Considerations for Adapting this System into Classrooms}

In a classroom environment, not every moment can be engaging. The goal is generally to minimize the long disengaged moments to keep the students' motivation high. So, it is vital to help teachers to analyze ML-based reports by combining multiple data sources. Based on our experience, we advise teachers to use ML-based engagement analysis when they already have some insights about students' characteristics, such as habits, likes, and dislikes. The ML models can only point out the engaged and disengaged moments based on some facial unit movements, body positions, and voice duration. This kind of guidance can be useful when students work together in a crowded classroom where the teacher cannot watch every group's behavior. However, in this scenario, the teacher still needs to take action, which can only be possible by re-evaluating the group dynamics and students' long-term behaviors.

\subsection{Limitations}
\label{subsec:limitations}

\textbf{Self-Evaluation of Engagement:} Self-evaluating the engagement level is challenging, as defining engagement in all aspects is still challenging for most researchers. Before the self-evaluation process, all participants completed a fifteen-minute training session that introduced the definition and dimensions of classroom engagement. Unlike the existing research, our system particularly focuses on capturing the multimodal aspect of classroom engagement. The labels of our dataset are ambiguous, and this ambiguous label represents a multi-dimensional complex term, ‘engagement.’  We collected data from group activities where students follow Youtube tutorials, hold group discussions, and complete hands-on, tangible activities. In this sense, our dataset is the only dataset that is collected in a controlled environment that represents the dynamic nature of learning.

\textbf{Representation Bias:} The collected data is from a small group, representing a very limited analysis of a small community. Considering these limitations, this data is limited for model training purposes but presents a unique dataset that is useful for exploratory analysis and testing.

\section{Conclusion}

Determining students' classroom engagement levels by giving attention to the right details is challenging, even for expert teachers. The multi-dimensional definition of engagement requires assessing students' affective, behavioral, and cognitive states. Developing a machine learning-based solution is a challenging task since the students' self-evaluation considers multiple aspects, but the recorded data presents only a few observable features. This paper presented a new dataset, its exploratory analysis and baseline deep learning architectures to accelerate the research in this challenging task. The dataset contains 26540 frames of audiovisual recordings with self-evaluation scores for each frame. Image frames, video slices, OpenFace and OpenPose feature vectors, audio formats, exploratory analysis scripts and deep learning model architectures are open-sourced to increase the accessibility of this dataset from both ML and education fields. Our image and video classification networks could achieve up to 68\% test accuracy on group-level engagement. We believe our comprehensive human-centered, open-source machine-learning pipeline can accelerate research in classroom engagement prediction. 

\section{Acknowledgements}

This project is supported by Koc University - Is Bank AI Center. We would like to thank for their genereous support. We would like to thank all participants to become a part of the dataset and our research.

\bibliographystyle{unsrt}  
\bibliography{sample} 

\end{document}